# Complexity, Stability Properties Of Mixed Games And *Dynamic Algorithms*, And "Learning" In The Sharing Economy.

Michael C. Nwogugu
Enugu 400007, Enugu State, Nigeria
Mcn2225@aol.com; mnwogugu91@gsb.columbia.edu
Skype: mcn1112
Phone: 234-909-606-8162

**Abstract.**
During 2005-2018, the "*Sharing Economy*" (which includes Airbnb; Apple; Alibaba, Uber; WeWork; Liquidspace, Lyft; Ebay; Didi Chuxing; Taxify; Amazon; HomeAway; etc.) blossomed across the world, triggered structural changes in industries and significantly affected international capital flows primarily by disobeying a wide variety of statutes laws in many countries and illegally reducing and changing the nature of competition in many industries often to the detriment of social welfare. Furthermore, Uber's and Airbnb's pricing systems are not efficient and can generate deadweight-losses and Regret. Other "sharing economy companies" ("SEOs") also face similar pricing, antitrust, deadweight-loss and Regret issues (eg. WeWork; Lyft; Taxify; LiquidSpace; ShareDesk; Homeaway; etc.). This article: i) develops new dynamic pricing models for the Sharing Economy companies; ii) derives some stability properties of *mixed games* and *dynamic algorithms* that are inherent in new dynamic pricing models introduced herein which eliminate antitrust liability for SEOs and also reduce deadweight losses, greed, Regret and GPS manipulation - that is, the new dynamic pricing models exhibit "Multi-Stability". The new dynamic pricing models have desirable properties such as *individual rationality*; *exact and approximate welfare maximization*; *strong budget balance*; *dominant-strategy incentive compatibility* of sellers and buyers; *pareto optimality* and *feasibility* – and thus, each model contravenes the *Myerson-Satterthwaite Impossibility Theorem*. The pricing models include multi-sided auctions wherein at any time *t*, several auctions can simultaneously occur, and the payoff functions of any buyer-seller pair in any auction depends on the bidding done by at least another buyer-seller pair either at the same time, or at a different time (this "*long-memory*" bias of buyers and sellers is new in the literature). More importantly the issues and models discussed herein have or can have significant implications for inequality, labor mobility and income dynamics (for employees of SEOs).

1. Introduction.
      The motivation for this article is varied. First, the "*Sharing Economy*" and the advent of digital currencies can be attributed to significant and rising consumer debt, inequality, low savings rates, internet penetration and the perceived inefficiencies of regulation. House-sharing companies (Airbnb, Homeaway), ride hailing/sharing companies (Uber; Lyft; Grab), App-stores (Google App Store; Apple IoS; Facebook app-store), office-space-sharing companies (WeWork; Liquidspace; ShareDesk), freelancer-sharing platforms (VWorker; Guru; ODesk; Elance, and Freelancer.com), shared-marketing platforms (Ebay; Amazon; Amazon-Flex; Alibaba), online filesharing platforms (Bittorent) and online information-sharing platforms (Instagram; Facebook) are among a group of internet based "*Sharing Economy*" companies and technologies (collectively, the "*Sharing Economy Organizations*" or "SEOs").
      Second, during 2010-2018, most SEOs thrived primarily by disobeying various laws in many countries - see Nwogugu (2017) which explains the many violations of statutes by SEOs including health (number of

occupants; sanitation; etc.); fire-protection (number of occupants; use of flammable materials; damage of fire doors; lack of knowledge of fire drills; etc.), housing, zoning, hotel, tax, securities, labor, taxi/transportation; insurance; antitrust; banking; electronic commerce; and consumer protection statutes. Government regulators often cannot afford to enforce those laws. There is abundant evidence that SEOs' past and current activities significantly harm cities and reduce social welfare in many economies.

Third, Uber's "*Surge Pricing*" is a form of dynamic pricing technology (dynamic pricing has been around for more than twenty five years). As of 2017, Uber's "Surge Pricing" was available only in high-demand periods and could result in prices that were ten times the normal prices. Mohlmann & Zalmanson (December 2017) and newspaper articles have noted that Uber's drivers knowingly colluded to increase "Surge Prices"[1]; and that Uber's customers can game the "Surge Pricing" system to reduce prices (eg. by crossing to the other side of the street and or by waiting for five minutes). Uber's drivers have been reported to have manipulated GPS readings (measurements of distance used for calculating Uber's fees) by using a third-party GPS app named Lockito[2].

Fourth, the SEOs' difficulties in developing and implementing efficient pricing mechanisms translates into, and amplifies other problems – such as antitrust violations; deadweight losses, Regret; etc..

Fifth, *dynamic algorithms* have not been addressed adequately in the literature, especially their stability properties. *Dynamic Algorithms* are a class of algorithms whose path or process changes or can change as state-variables change. On algorithms in general, see: Garcia, Berlanga, Molina & Davila (2004).

Sixth, mixed games have not been addressed adequately in the literature, especially their stability properties – mixed games are situations wherein two more games-types exist or can exist simultaneously and either continuously or discontinuously.

Seventh, the issues and models discussed herein have or can have significant implications for inequality, labor mobility and income dynamics (for employees of SEOs). Many such employees around the world depend on SEOs for all or part of their monthly income, which is related to their aspirations, social mobility and geographical mobility (eg. qualified drivers of Uber or Lyft ). However, its become clear that in addition to reducing Social welfare (increasing Noise; pollution; code violations; etc.) many SEOs also increase Inequality by underpaying their workers (through inefficient pricing algorithms).

By addressing these issues, this article contributes to the mechanism design, policy, dynamic algorithms, theoretical computer science; Labor Economics, Social Welfare and Complex Systems literatures.

---

[1] *See*: "*Uber Drivers Gang Up To Cause Surge Pricing, Research Says*". The Telegraph (UK). Cara McGoogan. August 2, 2017. http://www.telegraph.co.uk/technology/2017/08/02/uber-drivers-gang-cause-surge-pricing-research-says/. (stating that "Researchers at the University of Warwick found that Uber drivers in London and New York have been tricking the app into thinking there is a shortage of cars in order to raise surge prices. According to the study, drivers manipulate Uber's algorithm by logging out of the app at the same time, making it think that there is a shortage of cars. Uber raises its fare prices when there is a high demand for vehicles and a short supply of drivers available. Fares are known to increase during peak times such as rush hour, during public events and late at night. Surge pricing can boost the cost of rides to multiple times the normal rate. The study said drivers have been coordinating forced surge pricing, after interviews with drivers in London and New York, and research on online forums such as Uberpeople.net….…Separate research at Northeastern University (USA) has previously found passengers can game *surge pricing* with simple tricks such as waiting five minutes or crossing the road.……."). *See*: "*Uber Drivers Work Together To Create Price Surge And Charge Customers More, Researchers Find - Some Drivers Are Deliberately Going Offline In Unison So That Prices Surge And They Can Charge Customers More When They Log Back Into The App*". Ben Chapman, August 2, 2017. https://www.independent.co.uk/news/business/news/uber-drivers-work-together-price-surge-go-offline-charge-customers-more-game-app-supply-demand-a7872871.html.

[2] *See*: "*Uber Drivers In Lagos Are Using A Fake GPS App To Inflate Rider Fares*". Yemisi Adegoke. November 13, 2017. https://qz.com/1127853/uber-drivers-in-lagos-nigeria-use-fake-lockito-app-to-boost-fares/?utm_source=qzfb.

2. Existing Literature.

There is a growing academic and practitioner literature on the economics and regulation of SEOs[3] but all of the studies (eg. Edelman & Geradin (April 2016a;b); Edelman (2015; 2017)) didn't develop comprehensive pricing models that reduce SEOs' legal liabilities; and also didn't address some macroeconomic issues (such as international capital flows and structural changes which are intertwined with antitrust issues). As of 2017, many SEOs were defendants in court cases around the world for criminal and civil claims including antitrust violations, fraud, conspiracy, violations of criminal statutes; etc.. Edelman (June 2017) specifically advocated that Uber should be shut down because its business model is inherently illegal and its global legal liability far exceeds its total assets. Like Uber and as of 2017, many SEOs were technically insolvent because the statutes-of-limitations for their offenses had not expired (and the statute-of-limitations can be judicially extended in various ways) and their legal liability far exceeded their assets and equity. Malhotra & Van Alstyne (2014), Hellwig, Morhart, Girardin & Hauser (2015), Cusumano (2015), Cohen & Kietzmann (2014), Belk (2013; 2014) and Huefner (2015) discussed the sharing economy and applicable business models.

On *Deep Learning* and *Regret Minimization* in Mechanism Design, see: Feng, Narasimhan & Parkes (2018), Dutting, Feng, et. al. (2019) and Sandholm & Likhodedov (2015). On other Artificial Inteligence approaches, see: Rigas, Ramchurn & Bassiliades (2018), Zohar & Rosenschein (2008), Jain, Gujar, et. al. (2018), Aziz & Lev (2019), Peng, Al Chami, et. al. (2019), Gondran, Huguet, et. al. (2018), Chassaing, Duhamel & Lacomme (2016), Doshi, Gmytrasiewicz & Durfee (2020), and Mitchell (2006). Mohri & Medina (2014) and Nazerzadeh, Leme, et. al. (2016) analyzed learning algorithms. On algorithms in general, see: Garcia, Berlanga, Molina & Davila (2004).

Colini-Baldeschi, Goldberg, et. al. (Nov. 2017); Deng, Goldberg, Tang & Zhang (2014); Satterthwaite & Williams (1989), Dütting, Roughgarden & Talgam-Cohen (2014), Baranwal & Vidyarthi (2015), Wang, Chin & Yin (2011), Scalas, Rapallo & Radivojević (2017), Zhao (2012) and Zhang (2018) analyzed two-sided auctions. Wang & Wang (2015); Rao, Xiao, et. al. (2017); and Hara & Hato (2017) analyzed car-sharing auctions and transportation auctions. Liu, Zhu & Hu (2016) and Leung & Knottenbelt (2011) analyzed other types of auctions. Hill (2015), Techcrunch (2015), Krämer, Schmidt, Spann & Stich (2017), and Gibbs, Guttentag & Gretzel (2018) studied dynamic pricing in general and issues pertaining to Airbnb and Uber.

Banerjee, Johari & Riquelme (Nov. 2015) and Kung & Zhong (2017) discussed dynamic pricing models for ride-sharing platforms and the sharing economy respectively. On pricing in ride sharing/hailing networks, see: Afeche, Liu & Maglaras (2017), Afeche, Liu & Maglaras (2017), Banerjee, Freund & Lykouris (2016), Banerjee, Riquelme & Johari (2015), Bimpikis, Candogan & Saban (2016), Feng, Kong & Wang (2017), Nikzad (2018), and Ozkan & Ward (2016). On Surge-Pricing in ride sharing/hailing networks, see: Castillo, Knoepfle & Weyl (2017), Chan, Yom-Tov & Escobar (2014), Besbes, Castro & Lobel (2018), Mohlmann & Zalmanson (December 2017) and Cachon, Daniels & Lobel (2017). Srinivasan, Rajharhia, Radhakrishnan, Sharma & Khincha (2017); Thille,

---

[3] *See*: Edelman & Geradin (April 2016a); Edelman (2015); Edelman & Geradin (April 2016b); Edelman (June 2017); Dickerson & Hinds-Radix (April 2016). Estis & Lycoyannis (April 6, 2016); US Federal Trade commission (2016); Koopman, et al. (2015); and Eckhardt & Bardhi (2015).
*See*: "*Uber Scandals*". http://www.uberscandals.org/.
*See*: New York State Office of the Attorney General (Oct. 2014). "*Airbnb In The City*".
http://www.ag.ny.gov/pdfs/airbnb%20reportpdf.
*See: In the Matter of the Investigation by Eric T. Schneiderman, Attorney General of the State of New York (NYAG), of Uber Technologies, Inc.*. January 5, 2016.
*See*: "*Violent Massive Street Fighting In Jakarta Over Uber And Grab Taxi Services*".
www.eturbonews.com. March 22, 2016.
*See*: "*Nairobi's Taxi Drivers Turn To Violence To Halt Uber*". www.eturbonews.com
January 28, 2016.
*See*: "*Disruptive Innovation: Application Of Competition Law In The Sharing Economy In The Year Ahead*". Freshfields Bruckhaus Deringer, London, UK. http://antitrust.freshfields.com/disruptive-innovation.
*See*: European Commission – Press release: "*A European Agenda For The Collaborative Economy*".
http://europa.eu/rapid/press-release_IP-16-2001_en.htm.
*See*: Russo, F., "*Defining The Relevant Market In The Sharing Economy*". Available at:
https://policyreview.info/articles/analysis/defining-relevant-market-sharing-economy.

Cojocaru, et. al. (2013); Makhdoumi, Malekian & Ozdaglar (2017); and Nakhe (2017) developed dynamic pricing models for various markets.

Mashayekhy, Nejad & Grosu (2014) studied two-sided matching. Jacobs (2012) discussed coalgebras. Nissam (2007) summarized mechanism design. Feudel, Pisarchik & Showalter (2018); Wang, Xu & Lai (2018); Anzo-Hernández, Gilardi-Velázquez & Campos-Cantón (2018); and Kwasnioka (2018) analyzed multistability. Levina, Levin, et. al. (2006). Niels & Ten Kate (2000) analyzed Predatory Pricing.

2.1. LUPI And LUBA Are Not Applicable.

Ostling, Wang, Chou & Camerer (2011) and Zeng, Davis & Abbott (2007) analyzed the "*lowest unique positive integer*" game (LUPI) and related approaches. Bruss, Louchard & Ward (2009)[4]; Pigolotti, Bernhardsson, Juul, et. al. (2012); and Zhao, Chen & Wang (2013) analyzed inverse auctions that have unique minima (the "*lowest unique bid auction*" or "LUBA"). LUPI and LUBA are not directly applicable to the circumstances of many SEOs but are mentioned here as types of auctions. The analysis in Bruss, Louchard & Ward (2009) and in the literature on LUBA and LUPI isn't quite correct and the errors/omissions are as follows: i) the problem is not a two-sided problem in a mathematical or economics sense (lack of knowledge of the number of bidders and the distribution of bids does not make the problem two-sided); ii) the auction studied is not a two-sided auction because only one side (the buyer-side) bids; iii) the analogy to an "*urn problem*" is completely wrong and isn't applicable; iv) the process of Poissonization alone or together with de-Poissonization does not solve the problem of placing a winning bid because the process distorts data relationships (poissonization is based on the natural log scale whose non-linearity is unlikely to fit, or is not guaranteed to fit patterns in the auctions), and there is no guarantee that the bids will conform to any known distribution; and the number of bidders and the distribution of bids is unknown ex-ante; v) the bidding model should be based on, but does not consider the bidder's costs, budget constraints, mental states, propensity-to-substitute and opportunity costs.

2.2. Algorithm Collusion.

One issue that hasn't been sufficiently addressed in the literature (AI; Mechanism Design; Dynamic Pricing) is intentional and unintentional collusion by autonomous/intelligent pricing algorithms (of either the same company or of different/competing companies) which sometimes results in increases in final prices ("*Intentional Algorithm Collusion*" and "*Unintentional Algorithm Collusion*" respectively). As of 2019, most countries had not developed antitrust or consumer protection laws for such Collusion. Algorithm Collusion is relatively difficult to detect and prosecute. Also see the comments in: Calvano, Calzolari, Denicolò & Pastorello (2018); Ezrachi & Stucke (2015); Harrington (2018); Schwalbe (2018) and Kühn & Tadelis (2018).

In an empirical study with AI-based pricing algorithms in a controlled environment Calvano, Calzolari, et.al. (February 2019) noted that even relatively simple pricing algorithms learned and engaged in *Price Collusion* by trial and error, without any prior knowledge of their operating environment and without communicating with each another, and without being specifically designed or instructed to collude. *Algorithm Collusion* can be partly attributed to: i) the design, "learning-process" and updating of each pricing algorithm (how data is gathered and processed by each algorithm in each time period); ii) the relevance or "weights" assigned to each specific data source by each pricing algorithm; iii) the use of common sources of data by the pricing algorithms; iv) the knowledge/training of the designers of each pricing algorithms (designers who have the same knowledge and use the same approaches/models are more likely to cause *Algorithm Collusion*.

3. Complexity And Structural Changes.

The literature about the relationship between complexity and structural change is developed and includes Robert & Yoguel (2016); Cimoli, Pereima & Porcile (2016); Dosi & Virgillito (2017); Comim (2000); Heinrich & Dai (2016); and Brida, Anyul & Punzo (2003).

SEOs represent, have caused and continue to evolve major structural changes in several global industries (including housing, lodging, retailing/ecommerce, distribution, financial services and transportation) and their pricing models have been a distinctive feature. Such structural changes are often intertwined with antitrust issues. The pricing models of Uber, Lyft, Taxify, WeWork and Liquidspace represent structural changes in their respective industry segments. Uber's and Google's forays into driverless-cars marks another major change that can have significant implications for the manufacturing, transportation and internet/software sectors. The structural changes in these industries have been in the form of the following: i) the shift of transactions to the internet and changes in modes and timing of payment; ii) pricing models – which often affect competition, firm structure and industry structure; iii) antitrust issues; iv) technological advances, automation, reduction of labor and training of technology staff; v) industry consolidation through better and faster aggregation of individual suppliers and customers (as opposed to acquisitions of companies); v) increased *Network Effects* in industries; vi) significant reductions in the operating costs and marketing expenses of some companies (eg. SEOs and their hosts/merchants/drivers) some of which was achieved by disobeying laws; vii) low compliance (with regulations) by SEOs and their drivers/hosts/merchants; viii) major changes in, and evolution of the operations strategies of both SEOs and traditional incumbents (ie. brick-and-mortar hotels/resorts, retailing chains; taxi companies; distributors; manufacturers; etc.).

---

[4] *See*: http://en.wikipedia.org/wiki/Mobile_reverse_auction and http://en; and wikipedia.org/wiki/Unique_bid_auction.

The structural changes in these affected industries conform to the *Operations Strategy Model of Structural Change* (OSMSC) and other new theories of structural change that were introduced in Nwogugu (Revised 2013).

4. Some Antitrust Problems Of SEOs.

In 2016, the EU confirmed the antitrust liability of companies for anti-competitive conduct of their employees and agents[5]. Most of the SEOs discussed herein have two-sided platforms wherein their landlords/hosts/drivers/merchants are alleged to be independent contractors. Thus, such SEOs can be held liable for antitrust misconduct of their drivers/hosts/landlords/merchants. Furthermore, private equity firms have also been held liable for the antitrust misconduct of their portfolio company[6] – Nwogugu (2017) introduced the *Extended Duty Theory* which is different; and also explains the antitrust liabilities of SEOs.

See the comments: Dickerson & Hinds-Radix (April 2016); Cusumano (2015); Cohen & Kietzmann (2014); Nwogugu (2017); Niels & Ten Kate (2000); Mleczko (March 24, 2017); Markham (Oct. 2014); Khan (2017); Koopman, et. al. (2015); Heinrich & Dai (2016); Edelman & Geradin (April 2016b); Edelman (2015); and Eckhardt & Bardhi (2015).

5. New Dynamic Pricing Models That Reduce Antitrust Liability, Regret And Deadweight Losses; And Provides Driver Incentives And Customer-Incentives; And Reduces Manipulation Of GPS-Readings By Drivers.

This section introduces some new dynamic pricing models with embedded multi-sided auctions which are also dynamic algorithms. Several types of games (ie. matching games; assignment games; differential games; reverse-dutch auctions; and combinatorial auctions) and algorithms are simultaneously involved in the dynamic pricing models proposed herein, and most of the games are multi-stage non-cooperative games. However, the payoff-functions of each dynamic pricing model can account for each type of game involved. The models/solutions introduced herein are essentially dynamic algorithms and protocols whose processes/paths can change as state-variables change. The solutions/models can be built as online learning systems that gather, update and use information about state-variables. For taxi-SEOs, the state variables can include the number of participating drivers; ride fees; bidding patterns; allocation method; distances travelled; travel time; traffic conditions; types of customers; number of occupants in each car; types of drivers; types of cars; in-car conditions such as air-conditioning and no-smoking; etc.). For real estate SEOs such as Airbnb, the state variables can include the number of participating landlords and renters; rental fees; bidding patterns; allocation method; local market conditions; types of renters; number of occupants in each housing unit; types of landlords/sub-letters; types of housing units; in-house conditions such as air-conditioning and no-smoking; etc.).

---

[5] *See*: Thomas, C., DiStefano, G. & Jubrail, D. (Hogan Lovells) (Aug. 2016). *Antitrust Liability for Anti-competitive Behaviour by Employees and Contractors*. https://www.hoganlovells.com/en/blogs/focus-on-regulation/antitrust-liability-for-anti-competitive-behaviour-by-employees-and-contractors.

[6] *See*: Vinje, T. (Clifford Chance) (April 2014). *Private equity liability for antitrust fines*. https://www.cliffordchance.com/briefings/2014/04/private_equity_liabilityforantitrustfines.html. This article stated in part "……The European Commission has imposed a €37 million fine on Goldman Sachs (GS) for antitrust breaches committed by a portfolio company that was formerly owned by its private equity arm, GS Capital Partners. The fine was joint and several on GS and the portfolio company (Prysmian). It was imposed on the basis that GS exercised decisive influence over the portfolio company, though GS is not alleged to have participated in, been aware of or facilitated the alleged cartel in any way. ………."

The auctions introduced are *multi-sided auctions* because its highly probable that: i) at any time *t*, several auctions are occurring, and ii) the payoff functions of any buyer-seller pair in any auction depends on the bidding done by at least another buyer-seller pair either at the same time, or at a different time (this "Related-*Memory*" effect of buyers and sellers is new in the literature); iii) a seller (eg. a taxi driver) can simultaneously bid in different auctions for different contracts; and a buyer (eg. taxi customer) can open several auctions for bidding for similar contracts (eg. a taxi customer at Location-A can open 2-3 auctions for bidding for rides that start at Location-A or within 100 meters from Location-A).

As of 2017, Uber's "*Surge Pricing*" was available only in high-demand periods and could result in prices that were ten times the normal prices. Unlike Uber's "*Surge Pricing*" and traditional SEO pricing, the Type-1, Type-2, Type-3, Type-4 and Type-5 pricing schemes introduced herein and below can be available at any time, and they don't violate antitrust statutes (eg. price fixing; price discrimination; etc.); they reduce or eliminate *Regret* and *Deadweight Losses*; and eliminate manipulation of GPS readings by drivers; and can improve social welfare.

The new dynamic-pricing models introduced herein differ from both LUBA and LUPI in the following ways: i) only the lowest bid is used, and it doesn't have to be a unique-bid, and where two persons bid the same lowest amount, the BBB App will select the bidder with the lower estimated arrival time and or the higher quality score; ii) the drivers and customers don't have to pay any fee to participate in the bidding, but may pay a fee if they don't bid when selected; iii) the auction is two-sided because for each ride, both the drivers and the customer bids; iv) the auction is "*cross-contingent*" because for each ride, the drivers' bids are or can be affected by other available auctions for other rides in the same network, and the customers' bid is affected by other customers' bids in the network; v) the bidding processes is part of a matching process; vi) the bidding process can reduce deadweight losses.

Thus, the auctions introduced herein are two new classes of auctions which are as follows:
i) The "*two-sided cross-contingent auctions*".
ii) The "*double-bid two-sided auction*" – wherein both sides of a two-sided auction simultaneously bid.

Each of the Type-1, Type-2, Type-3, Type-4 and Type-5 pricing schemes herein and below are a new type of dynamic pricing henceforth referred to as "*Two-sided State-Contingent dynamic pricing*". The pricing system is two-sided because both sides of the transaction (the drivers and the customers) can participate in the price-setting process; and at any time, more than one customer can use the system, and a customer's bidding and use of BBB's vehicle affects or can affect other customer's prices and bidding. The customers in each city essentially compete for a finite number of BBB vehicles; while drivers compete for a finite number of customers). The system is state-contingent because each party in both sides of the transaction (drivers and the customer) can choose more than one "state". The pricing scheme can also include a "*Fulfillment Guarantee*" wherein the booked customer must appear and take the ride within a specific number of minutes after the driver arrived at the designated location, and if the customer defaults, he/she will pay a fee to BBB, part of which will be paid to the driver. Where the *Arrival-Guarantee* and or *Fulfillment Guarantee* is included in the pricing scheme, each of the Type-1, Type-2, Type-3 and Type-4 pricing schemes are a new type of dynamic pricing henceforth referred to as "*Two-sided State-Contingent Time-Contingent dynamic pricing*". The system is time-contingent because at least one party on each side of the transaction (driver or customer) has a time-based obligation that directly affects the pricing mechanism.

5.1. Common Elements Of The New Dynamic Pricing Schemes – Anti-Collusion; Privacy; Preferences; Incentives; Etc..

The following are some common factors among the Type-1, Type-2, Type-3, Type-4 and Type-5 dynamic pricing schemes:
> i) Each of the new pricing mechanisms can be a combinatorial auction because each customer is bidding for a group of discrete items such as arrival-time (of the driver), the taxi-ride, amenities in the taxi-vehicle; the driver's and vehicle's compliance with applicable regulations. Conversely, each driver is bidding for a group of discrete items such as arrival-time (of the customer), the taxi-ride, and fees to be paid by the customer.
> ii) Each of the new pricing mechanisms are multi-sided auctions because for each "contract" (eg. a taxi-ride), several providers (drivers) can simultaneously bid for the contract and other un-related contracts; and the customer can also create new related and un-related contracts.
> iii) Each of the new pricing mechanisms are *matching games; assignment games* and types of *reverse-dutch auctions*.

iv) In some cities, taxi drivers gather/line-up at specific locations to chat or pickup customers and can collude to set prices and or the timing of rides. In other instances, drivers that are not in proximity can also collude to switch off their phones at certain times or not to accept rides. For Type-3 and Type-4 pricing schemes, drivers that collude to engage in price-fixing or price discrimination can be discovered and will be penalized. For example, such colluding drivers can be discovered by: 1) by checking for loose groups of drivers that bid from one neighborhood wherein their bids are within a 10%-/+ band of the median bid for each customer for a specific number of customers within a time frame; 2) checking for drivers that switch off their cellphones at or around the same time on more than two occasions.

v) In dynamic pricing systems for taxi-companies, setting the driver's "*Preferences*" to "automatic states" can drastically reduce the distractions and accidents that occur when drivers use their cellphones while driving.

vi) For each of the Type-1, Type-2, Type-3, Type-4 and Type-5 pricing schemes, each customer's bid will not be disclosed to any other customer or any driver that is not among the 3-15 selected drivers; and each driver's bid for any customer will not be disclosed to any other driver or customer; and the customer's identity will not be disclosed to any driver, and the bidding drivers' identity will not be disclosed to any customer, and only the winning/assigned driver will be introduced to the customer after the customer pays for the ride.

vii) In cities/towns that have adequate GIS data, and for each of the Type-1, Type-2, Type-3, Type-4 and Type-5 pricing schemes, the entire price-setting process can be 100% automated (no human intervention) in BBB's App and can take less than five minutes.

viii) Each of the Type-1, Type-2, Type-3, Type-4 and Type-5 pricing schemes can include an "*Arrival Guarantee*" wherein at the time of bidding, each driver also posts a guaranteed time-of-arrival at the customer's location and if the driver defaults, the customer will get a 15%-35% price discount which will be deducted from the driver's pay. Similarly, if the customer doesn't arrive within ten minutes of the scheduled time, then the customer will be charged a specific fee based on his/her subsequent arrival.

ix) Each of the Type-1, Type-2, Type-3, Type-4 and Type-5 pricing schemes can include an "*Incentive Fee*" wherein at the time of bidding, each driver will be notified of a specific incentive that is payable to him/her solely for completing the ride within a specific period of time.

x) In each of the Type-1, Type-2, Type-3, Type-4 and Type-5 pricing schemes, the App will be designed so that drivers wont be able to manipulate GPS readings: 1) BBB will install its own GPS sensor in the vehicle (rather than using the GPS in the driver's phone); 2) all GPS readings will be controlled by BBB's headquarters and not by the driver's phone (even if the GSP in the driver's phone is being used); 3) the system will note the vehicle's location 4-5 times during the ride in other to measure distance; 3) at inception of the ride the App/system will calculate the distance to be travelled and the basic fare before the bidding begins.

xi) The Related-Memory Effect of buyers and sellers – see above.

xii) Each of the Type-1, Type-2, Type-3, Type-4 and Type-5 pricing schemes can be used for "car-pooling", "flexible car-pooling, "van-pooling" and "P2P ride-sharing" models.

The concepts of *Algorithmic Mechanism Design* and *Complexity Theory* are applicable; and BBB's prices can be set as follows. On algorithms in general, see: Garcia, Berlanga, Molina & Davila (2004).

5.2. Type-1 Pricing.

**Theorem-1A**: *For All Dynamic Pricing Systems That Are Both "Open" Dynamical Systems And "Open" Dynamic Games, There Is At Least One Dynamic Pricing Mechanism (Type-1 Pricing) That Can Reduce Deadweight Losses And Antitrust Liability, And Involves "two-sided cross-contingent auctions" And "Two-sided State-Contingent dynamic pricing".*

*Proof*: "Open" refers to the fact that both the number of participants and the number of client-provider combinations in the pricing mechanism are potentially unlimited, and while the pricing mechanism has some "*invariants*", it changes constantly. The dynamic pricing mechanism is as follows. In *Type-1 pricing*, once the customer books a ride, the BBB App will provide a "Standard Price" ($\Omega_{cs}$) that is the actual meter rates used by licensed taxis in the city (or if there isn't any standard meter rate, then BBB's own publicized meter rate which should be less than the black-car taxi rate), and the customer will "bid"/specify the minimum and maximum prices that he/she is willing to pay for the ride (hereafter, the "*Customer Range*" or the set $C_r$; and $C_n$ is a price within the set $C_r$ that satisfies the condition $C_n > (\Omega_{cb}, 0)$). The information will be automatically relayed to the 3-15 BBB cars that are nearest to the customer, all of whom are required to respond by bidding within minutes. The *Base Price* ($\Omega_{cb}$), is the minimum price required for BBB to earn a profit from the ride less any temporary customer-specific or ride-specific losses that BBB will absorb in order to build market share or brand equity. For each such ride, each BBB driver can pre-set his/her "preferences" in the BBB App: i) to manually "bid" a specific price only if such price exceeds the Base Price (state "$S_{d1}$"); or ii) to automatically bid the mid-point of the price range offered by the customer, only if such price exceeds the Base Price (state "$S_{d2}$"); or iii) to not submit any bid (state "$S_{d3}$") (and collectively, the price bid by driver $i$ in the set of drivers $1........n$ is $R_{id}$; and the lowest price among all bids by all drivers is $\Omega_{cd}$). Each driver can simultaneously bid for many customers, but any driver that wins a bid cannot bid for any other ride that occurs within one hour before or after his/her winning bid. The customer can revise his/her *Customer Range* only twice, and only upwards and within ten minutes after his/her first bid (states "$S_{c1}$", "$S_{c2}$" and "$S_{c3}$" respectively). After each such revision, each participating driver can revise his/her bid only once. The BBB App will automatically assign a primary-driver (with the lowest bid or a combination of one of the lowest bids and lowest arrival-times) and a secondary driver (with the next highest bid) to the customer to provide the ride. All payments will be made through the BBB app. The model reduces *deadweight losses* because it dynamically reduces both *consumer-surplus* and *producer-surplus*. For each ride, BBB will pay the driver a fee equal to the lesser of: i) the average percentage of revenues that black-car taxi companies pay their drivers in that city; or ii) X%-Y% (eg. 75%-90%) of the fees paid by the customer less applicable taxes and less any fees payable to third-party car-owners – the variables include commercial insurance fees, road-worthiness testing costs; traffic conditions in the city; the type/class of car used; the customer's gratuities; and the driver's "quality-scores"; etc. (collectively, the fee is "$\psi_{dd}$"). BBB's variable costs and allocated fixed costs for the ride are $\beta_v$ and $\beta_f$ respectively. BBB's opportunity cost is the net profit ("$\psi_{ob}$" – the "net" incremental profit over and above that of the matched-ride) from winning and servicing another customer with a probability of $\beta_{ob}$; and the probability $\beta_{lb}$ that another customer with a larger fare exists. The driver's own variable costs and allocated fixed costs for the ride are $D_v$ and $D_f$ respectively. The driver's opportunity cost is the net profit ("$\psi_{od}$" – the "net" incremental profit over and above that of the matched-ride) from bidding for, and winning and servicing another customer with a probability of $\beta_{od}$; and the probability $\beta_{ld}$ that another customer with a larger fare exists. The probability that at least one driver will bid for a customer-ride is $\beta_{id}$. Note that in most cases, and for each driver-customer pair, $\beta_{ob}$ and $\beta_{od}$ will be different and based on different sets of customers. Both probability-adjusted opportunity costs and the two-sided bidding/price-setting protocols incorporate dynamic *Regret Minimization* into the dynamic pricing model (ie. the reduction in *Regret* can be significant and varies with changes in state variables). The probability that driver $i$'s bid wins is $\beta_{idw}$ (and the probabilities for drivers $1........n$ are $\beta_{1dw}......\beta_{ndw}$). Generally, $\Omega_{cb} < \Omega_{cd} < \Omega_{cs} < C_n$. ∎

**Theorem-1B**: *For All Prices That Are Real Numbers, There Is Least One Equilibrium Payoff That Satisfies The Conditions Of Type-1 Pricing And The Elements Of The Mixed Games (eg. multi-sided auctions; matching games; assignment games; differential games; reverse-dutch auctions; multi-stability; and combinatorial auctions) And Dynamic Algorithms Involved; And Can Reduce Deadweight Losses And Antitrust Liability.*
***Proof***:
    BBB's objective function and payoff for the ride is:
$P_{o1} = \text{Max} [(\min\{C_n; \text{Max}(0;(\Omega_{cd} | \text{Min}(S_{d1},S_{d2},S_{d3})); \Omega_{cb}))\} - (\psi_{ob}*(\beta_{ob} | \beta_{lb})) - (\psi_{dd}+\beta_f+\beta_v))* \beta_{id})]$; *iff* $C_n > (\Omega_{cb}; \Omega_{cd})$; or

$P_{o2} = \text{Max} [((\text{Max}(0;(\Omega_{cd} | \text{Min}(S_{d1},S_{d2},S_{d3})); \Omega_{cb}) - (\psi_{ob}*(\beta_{ob} | \beta_{lb})) - (\psi_{dd}+\beta_f+\beta_v))*\beta_{id})]$; *iff* $C_n < (\Omega_{cb}; \Omega_{cd})$

The driver's simple payoff function is:
$P_s = \text{Max}[\{(\beta_{id} | \beta_{idw})*(\psi_{dd} - D_v - D_f - (\psi_{od}*(\beta_{od} | \beta_{ld})))\}; 0]$; where $(0 < \psi_{dd}) \in \Omega_{cd}$

    These formulas can also answer the question of when a consumer will join a sharing system – and generally and with limited consideration of the customers' non-monetary utility from expedited service and or convenience, that becomes more probable as $[\text{Max}(P_{o1}, P_{o2}) + P_s] \to -\infty$. ∎

5.3. Type-2 Pricing.

**Theorem-2A**: *For All Dynamic Pricing Systems That Are Both "Open" Dynamical Systems And "Open" Dynamic Games, There Is At Least One Dynamic Pricing Mechanism (Type-2 Pricing) That Can Reduce Deadweight Losses And Antitrust Liability, And Involves "two-sided cross-contingent auctions" And "Two-sided State-Contingent dynamic pricing".*

*Proof*: "Open" refers to the fact that both the number of participants and the number of client-provider combinations in the pricing mechanism are potentially unlimited, and while the pricing mechanism has some "*invariants*", it changes constantly. The dynamic pricing mechanism is as follows. In *Type-2 pricing*, once the customer books a ride, the BBB App will provide "estimated prices" from 3-7 drivers that are nearest to such customer and the prices will be capped at 70%-85% of the official meter rates used by licensed taxis in that city or a publicized BBB meter rate (which should be less than the local average black-car taxi rate) if there isn't any official taxi meter rate (collectively, the "Standard Price" or $\Omega_{cs}$). The Standard Price will be greater than a Base Price (the Base Price, $\Omega_{cb}$, is the minimum price required for BBB to earn a profit from the ride). Each driver is required to respond within minutes by: i) automatically setting his/her "preferences" in the BBB App to bid at the "*Standard Price*" offered by BBB to the customer (state "$S_{d1}$"), or ii) automatically setting his/her "preferences" in the BBB App to manually bid at prices between the Base Price and the Standard Price (state "$S_{d2}$"); iii) automatically setting his/her "preferences" in the BBB App not to bid (state "$S_{d3}$") (and collectively, the resulting price for driver *i* in the set of drivers $1……n$, is $R_{id}$; and the lowest price among all bids by all drivers is $\Omega_{cd}$). Each driver can simultaneously bid for many customers, but any driver that wins a bid cannot bid for any other ride that occurs within a specific time (eg. 30 minutes) before or after his/her winning bid. All drivers must inform BBB about when they are off-duty and cannot return to duty unless a specific time elapses; and any driver that doesn't bid for a customer will pay a small fee to BBB. The customer can revise the Standard Price only once and only downwards but above the Base Price and within ten minutes after the drivers' bids are sent to him/her (states "$S_{c1}$" and "$S_{c2}$" respectively). The BBB App will automatically assign a primary-driver (with the lowest bid or a combination of one of the lowest bids and lowest arrival-times) and a secondary driver (with the next highest bid – just in case the primary driver doesn't arrive on time) to the customer to provide the ride. Independent unaffiliated commercial drivers whose vehicles have commercial plates can also bid for Type-2 customers by downloading the BBB App and bidding at prices between the Base Price and the Standard Price. All payments will be made through the BBB app. BBB will pay the driver a fee equal to the lesser of: i) the average percentage of revenues that black-car taxi companies pay their drivers in that city less applicable taxes and less any fees payable to third-party car-owners; or ii) X%-Y% (eg. 75%-90%) of the fees paid by the customer less applicable taxes and less any fees payable to third-party car-owners – the variables include commercial insurance fees, road-worthiness testing costs; traffic conditions in the city; the type/class of car used; the customer's gratuities; and the driver's "quality-scores"; etc. (collectively, "$\psi_{dd}$"). BBB's variable costs and allocated fixed costs for the ride are $B_v$ and $B_f$ respectively. BBB's opportunity cost is the net profit ("$\psi_{ob}$" – the "net" incremental profit over and above that of the matched-ride) from winning and servicing another customer with a probability of $\beta_{ob}$; and the probability $\beta_{lb}$ that another customer with a larger fare exists. The driver's own variable costs and allocated fixed costs for the ride are $D_v$ and $D_f$ respectively. The driver's opportunity cost is the net profit ("$\psi_{od}$" – the "net" incremental profit over and above that of the matched-ride) from bidding for, and winning and servicing another customer with a probability of $\beta_{od}$; and the probability $\beta_{ld}$ that another customer with a larger fare exists. Both probability-adjusted opportunity costs and the two-sided bidding/price-setting protocols incorporate elements of dynamic *Regret Minimization* into the dynamic pricing model (ie. the reduction in *Regret* can be significant and varies with changes in state variables). The probability that at least one driver will bid for a customer-ride is $\beta_{id}$. Note that in most cases, and for each driver-customer pair, $\beta_{ob}$ and $\beta_{od}$ will be different and based on different sets of customers. The probability that driver *i*'s bid wins is $\beta_{idw}$ (and the probabilities for drivers $1…….n$ are $\beta_{1dw}……\beta_{ndw}$). ∎

**Theorem-2B**: *For All Prices That Are Real Numbers, There Is Least One Equilibrium Payoff That Satisfies The Conditions Of Type-2 Pricing And The Elements Of The Mixed Games (eg. multi-sided auctions; matching games; assignment games; differential games; reverse-dutch auctions; multi-stability; and combinatorial auctions) And Dynamic Algorithms Involved; And Can Reduce Deadweight Losses And Antitrust Liability.*
*Proof*:
    BBB's objective function and payoff for the ride is:
$P_o = \text{Max } [(\min\{\Omega_{cs}; \text{Max}(\Omega_{cb}; 0; (\Omega_{cd} | \text{Min}(S_{d1},S_{d2},S_{d3})))\} - (\psi_{ob}*(\beta_{ob} | \beta_{lb})) - (\psi_{dd}+\beta_f+\beta_v))*\beta_{id}]$

The driver's payoff is:
$P_s = \text{Max}[\{(\hat{\beta}_{id} | \hat{\beta}_{idw})*(\psi_{dd} - D_v - D_f - (\psi_{od}*(\hat{\beta}_{od} | \hat{\beta}_{ld})))\}; 0]$; where $(0 < \psi_{dd}) \in \Omega_{cd}$

These formulas can also answer the question of when a consumer will join a sharing system – and generally and with limited consideration of the customers' non-monetary utility from expedited service and or convenience, that becomes more probable as $(P_o + P_s) \to -\infty$.

The dynamic pricing model reduces *deadweight losses* because it dynamically reduces both *consumer-surplus* and *supplier-surplus* (ie. the reductions vary with the bidding patterns and other state variables). ∎

5.4. Type-3 Pricing.

**Theorem-3A**: *For All Dynamic Pricing Systems That Are Both "Open" Dynamical Systems And "Open" Dynamic Games And Wherein Priority-Of-Service Is Important, There Is At Least One Dynamic Pricing Mechanism (Type-3 Pricing) That Can Reduce Deadweight Losses And Antitrust Liability, And Involves "two-sided cross-contingent auctions" And "Two-sided State-Contingent dynamic pricing".*

*Proof*: "Open" refers to the fact that both the number of participants and the number of client-provider combinations in the pricing mechanism are potentially unlimited, and while the pricing mechanism has some "*invariants*", it changes constantly. The dynamic pricing mechanism is as follows. In *Type-3 "Priority1 Pricing"* (which targets customers that volunteer to pay more in order to get priority service at any time), once a customer books a ride, BBB will offer a "*Standard Price*" ($\Omega_{cs}$) which will not exceed the taxi meter-rate used in that city (or if there isn't any official taxi meter rate, then a publicized BBB meter rate which should be less than the local average black-car taxi rate) but will be greater than a Base Price. The *Base Price* ($\Omega_{cb}$) is the minimum price required for BBB to earn a profit from the ride, minus any ride-specific losses that that BBB is willing to absorb solely for customer retention or brand building. The customer will have the opportunity (but is not required) to "bid" the highest price that he/she is willing to pay for that ride (a "*Priority1-Price*" or $\Omega_p$) which must exceed the *Standard Price*. The customer can revise the *Priority1-Price* only twice and only upwards and within ten minutes after his/her first bid (states "$S_{c1}$", "$S_{c2}$" and "$S_{c3}$" respectively). After each such revision, each participating driver can change his/her bid only once. The BBB App will then automatically relay the *Priority1-Price* (and any revisions) to 3-15 drivers that are nearest and have the lowest arrival-time to the customer's location. Each of the 3-15 selected drivers can respond by setting their "preferences" (in the BBB App) to only one of five states: i) automatically accept the Priority Price (state "$S_{d1}$"); or ii) to bid a percentage of the *Priority1-Price* (eg. 70%-99% of the *Priority1-Price*) that exceeds the Standard Price (state "$S_{d2}$"); or iii) to accept the Standard Price (state "$S_{d3}$"); or iv) to manually bid at prices between the Priority Price and the Base Price (state "$S_{d4}$"); or v) not to bid (state "$S_{d5}$") (and collectively, the price bid by driver *i* in the set of drivers 1……..n is $\Omega_{id}$; and the lowest price among all bids by all drivers is $\Omega_{cd}$ – also the low price bid by the selected driver with the lowest price-arrival combination). The BBB App will automatically assign a primary-driver (with the lowest bid or a combination of one of the lowest bids and lowest arrival-times) and a secondary driver (with the next higher bid) to the customer to provide the ride. Each driver can simultaneously bid for many customers, but any driver that wins a bid cannot bid for any other ride that occurs within a specific time (eg. thirty minutes) before or after his/her assigned ride. All drivers must inform BBB about when they are off-duty and cannot return to duty unless a specific time elapses; and any driver that doesn't bid for a customer will pay a small fee to BBB; and any driver that is selected and that bids, must deliver the ride. Independent unaffiliated commercial drivers whose vehicles have commercial plates can also bid for Type-3 customers by downloading the BBB App and bidding below the Priority Price. The customer will be notified of a final price before he/she makes payment. All payments will be made through the BBB app. BBB will pay drivers a fee equal to the lesser of: i) the average percentage of revenues that black-car taxi companies pay their drivers in that city less applicable taxes and less any fees payable to third-party car-owners; or ii) X%-Y% (eg. 75%-90%) of the fees paid by the customer less applicable taxes and less any fees payable to third-party car-owners – the variables include commercial insurance fees, road-worthiness testing costs; traffic conditions in the city; the type/class of car used; the customer's gratuities; and the driver's "quality-scores"; etc. (collectively, the fee is "$\psi_{dd}$"). BBB's variable costs and allocated fixed costs for the ride are $\beta_v$ and $\beta_f$ respectively. BBB's opportunity cost is the net profit ("$\psi_{ob}$" – the "net" incremental profit over and above that of the matched-ride) from winning and servicing another customer with a probability of $β_{ob}$; and the probability $β_{lb}$ that another customer with a larger fare exists. The driver's own variable costs and allocated fixed costs for the ride are $D_v$ and $D_f$ respectively. The driver's opportunity cost is the net profit ("$ψ_{od}$" – the "net" incremental profit over and above that of the matched-ride) from bidding for, and winning and servicing another customer with a probability of $β_{od}$; and the probability $β_{ld}$ that another customer with a larger fare exists. Both probability-adjusted opportunity costs and the two-sided bidding/price-setting protocols incorporate dynamic *Regret Minimization* into the dynamic pricing model (ie. the reduction in *Regret* can be significant and varies with changes in state variables). See Nwogugu (2006). The probability that at least one driver will bid is $β_{id}$. Note that in most cases, and for each driver-customer pair, $β_{ob}$ and $β_{od}$ will be different and based on different sets of customers. The probability that driver *i*'s bid wins is $β_{dw}$. Generally, $Ω_{cb} < Ω_{cd} < Ω_{cs}$. ∎

**Theorem-3B**: *For All Prices That Are Real Numbers, There Is Least One Equilibrium Payoff That Satisfies The Conditions Of Type-3 Pricing And The Elements Of The Mixed Games (eg. multi-sided auctions; matching games; assignment games; reverse-dutch auctions; multi-stability; and combinatorial auctions) And Dynamic Algorithms Involved; And Can Reduce Deadweight Losses And Antitrust Liability.*
*Proof*:
BBB's objective function and payoff is:
$P_o$ = Max [(min{$Ω_p$; Max(($Ω_{cd}$ | Min($S_{d1}$;$S_{d2}$;$S_{d3}$;$S_{d4}$;$S_{d5}$)); $Ω_{cs}$; 0; $Ω_{cb}$)} -($ψ_{ob}$*($β_{ob}$ | $β_{lb}$))-($ψ_{dd}$+$β_f$+$β_v$))*$β_{id}$)].

The driver's payoff function is:
$P_s$ = Max[{($β_{id}$ | $β_{idw}$)*($ψ_{dd}$- $D_v$ -$D_f$-($ψ_{od}$*($β_{od}$ | $β_{ld}$)))}; 0]; where (0<$ψ_{dd}$) ϵ $Ω_{cd}$.

These formulas can also answer the question of when a consumer will join a sharing system – and generally and with limited consideration of the customers' non-monetary utility from expedited service and or convenience, that becomes more probable as ($P_o$+$P_s$) → -∞.
The model reduces *deadweight losses* because it dynamically reduces both *consumer-surplus* and *supplier-surplus* (ie. the reductions vary with the bidding patterns and other state variables). ∎

5.5. Type-4 Pricing.

**Theorem-4A**: *For All Dynamic Pricing Systems That Are Both "Open" Dynamical Systems And "Open" Dynamic Games And Targets "Price-And-Value" Sensitive Customers, There Is At Least One Dynamic Pricing Mechanism (Type-4 Pricing) That Can Reduce Deadweight Losses And Antitrust Liability, And Involves "two-sided cross-contingent auctions" And "Two-sided State-Contingent dynamic pricing".*

*Proof*: "Open" refers to the fact that both the number of participants and the number of client-provider combinations in the pricing mechanism are potentially unlimited, and while the pricing mechanism has some "*invariants*", it changes constantly. The dynamic pricing mechanism is as follows. In *Type-4 "Flex Pricing"* (which targets price/value sensitive customers), once a customer books a ride, BBB will offer a "*Standard Price*" ($Ω_{cs}$) which will not exceed the taxi meter-rate used in that city (or a publicized BBB meter rate if there isn't any official taxi meter rate) but will be greater than a Base Price (the *Base Price* $Ω_{cb}$, is the minimum price required for BBB to earn a profit from the ride, minus any losses that that BBB is willing to absorb solely for customer retention or brand building). The customer will be required to "bid" a price $C_n$ that is within a "*Flex Range*" ("$C_r$"; a set which is a 20%-/+ band above and below the *Standard Price*). This "*Flex Price*" $C_n$ is a price within the set $C_r$ that satisfies the condition $C_n$>($Ω_{cb}$,0). The BBB App will then automatically relay the *Flex-Price* to 3-15 drivers that are nearest to, and have the lowest arrival-time to the customer's location. Each of the 3-15 selected drivers can respond by setting their "preferences" (in the BBB App) to only one of five states: i) automatically bid/accept the Flex Price (state "$S_{d1}$"); or ii) to bid at the Standard Price (state "$S_{d2}$"); or iii) to bid a pre-set percentage of the Standard Price that is within the Flex Range (state "$S_{d3}$"); or iv) to manually bid at prices within the Flex Range (state "$S_{d4}$"); or v) not to bid (state "$S_{d5}$") (and collectively, the price bid by driver *i* in the set of drivers 1……..*n* is $Ω_{id}$; and the lowest price among all bids by all drivers is $Ω_{cd}$). BBB will assign a primary-driver (a driver with the lowest bid or a combination of one of the lowest bids and lowest arrival-times) and a secondary driver (with the next highest bid) to the customer to provide the ride. Each driver can simultaneously bid for many customers, but any driver that wins a bid cannot bid for any other ride that occurs within a specific time (eg. 30 minutes) before or after his/her winning bid. All drivers must inform BBB about when they are off-duty and cannot return to duty unless a specific time elapses; and any driver that doesn't bid for a customer will pay a small fee to BBB. Independent unaffiliated commercial drivers whose vehicles have commercial plates can also bid for Type-3 customers by downloading the BBB App and bidding below the Priority Price. The customer will be notified of a final price before he/she makes payment. All payments will be made through the BBB app. BBB will pay drivers a fee equal to the lesser of: i) the average percentage of revenues that black-car taxi companies pay their drivers in that city less applicable taxes and less any fees payable to third-party car-owners; or ii) X%-Y% (eg. 75%-90%) of the fees paid by the customer less applicable taxes and less any fees payable to third-party car-owners – the variables include commercial insurance fees, road-worthiness testing costs; traffic conditions in the city; the type/class of car used; the customer's gratuities; and the driver's "quality-scores"; etc. (collectively, the fee is "$\psi_{dd}$"). BBB's variable costs and allocated fixed costs for the ride are $\beta_v$ and $\beta_f$ respectively. BBB's opportunity cost is the net profit ("$\psi_{ob}$" – the "net" incremental profit over and above that of the matched-ride) from winning and servicing another customer with a probability of $\beta_{ob}$; and the probability $\beta_{lb}$ that another customer with a larger fare exists. The driver's own variable costs and allocated fixed costs for the ride are $D_v$ and $D_f$ respectively. The driver's opportunity cost is the net profit ("$\psi_{od}$" – the "net" incremental profit over and above that of the matched-ride) from bidding for, and winning and servicing another customer with a probability of $\beta_{od}$; and the probability $\beta_{ld}$ that another customer with a larger fare exists. Both probability-adjusted opportunity costs and the two-sided bidding/price-setting protocols incorporate elements of dynamic *Regret Minimization* into the dynamic pricing model (ie. the reduction in *Regret* can be significant and varies with changes in state variables). The probability that at least one driver will bid is $\beta_{id}$. Note that in most cases, and for each driver-customer pair, $\beta_{ob}$ and $\beta_{od}$ will be different and based on different sets of customers. The probability that driver *i*'s bid wins is $\beta_{idw}$ (and the probabilities for drivers 1…….n are $\beta_{1dw}$……$\beta_{ndw}$). Generally, its hoped that $\Omega_{cb} < \Omega_{cd} < \Omega_{cs} < C_n$. ■

**Theorem-4B**: *For All Prices That Are Real Numbers, There Is Least One Equilibrium Payoff That Satisfies The Conditions Of Type-4 Pricing And The Elements Of The Mixed Games (eg. multi-sided auctions; matching games; assignment games; differential games; reverse-dutch auctions; multi-stability; and combinatorial auctions) And Dynamic Algorithms Involved; And Reduces Deadweight Losses And Antitrust Liability.*
*Proof*:
BBB's objective function and payoff for the ride is:

$P_o = Max [((\{Max((R_{cd} | Min(S_{d1};S_{d2};S_{d3};S_{d4};S_{d5})); 0; \Omega_{cb}) \epsilon C_n\} - (\psi_{ob}*(\beta_{ob} | \beta_{lb})) - (\psi_{dd}+\beta_f+\beta_v))*\beta_{id})].$

The driver's simple payoff function for the ride is:
$P_s = Max[\{(\beta_{id} | \beta_{idw})*(\psi_{dd} - D_v - D_f - (\psi_{od}*(\beta_{od} | \beta_{ld})))\}; 0]$; where $(0<\psi_{dd}) \epsilon \Omega_{cd}$.

These formulas can also answer the question of when a consumer will join a sharing system – and generally and with limited consideration of the customers' non-monetary utility from expedited service and or convenience, that becomes more probable as $(P_o+P_s) \rightarrow -\infty$.
The model reduces *deadweight losses* because it dynamically reduces both *consumer-surplus* and *supplier-surplus* (ie. the reductions vary with the bidding patterns and other state variables). ■

5.6. Type-5 Pricing.

**Theorem-5A**: *For All Dynamic Pricing Systems That Are Both "Open" Dynamical Systems And "Open" Dynamic Games And Wherein Priority-Of-Service Is Important, There Is At Least One Dynamic Pricing Mechanism (Type-5 Pricing) That Can Reduce Deadweight Losses And Antitrust Liability, And Involves "two-sided cross-contingent auctions" And "Two-sided State-Contingent dynamic pricing".*

*Proof*: "Open" refers to the fact that both the number of participants and the number of client-provider combinations in the pricing mechanism are potentially unlimited, and while the pricing mechanism has some "*invariants*", it changes constantly. The dynamic pricing mechanism is as follows. In *Type-5 "Priority2 Pricing"* (which targets customers that volunteer to pay more in order to get priority service at any time), once a customer books a ride, BBB will offer a "*Standard Price*" ($\Omega_{cs}$) which will not exceed the taxi meter-rate used in that city (or if there isn't any official taxi meter rate, then a publicized BBB meter rate which should be less than the local average black-car taxi rate) but will be greater than a *Base Price*. The *Base Price* ($\Omega_{cb}$) is the minimum price required for BBB to earn a profit from the ride, minus any ride-specific losses that that BBB is willing to absorb solely for customer retention or brand building. The customer will be required to "bid" the lowest price that he/she is willing to pay for that ride (a "*Priority2 Price*" or $\Omega_p$) which must exceed the *Base Price*. The customer can revise the *Priority2-Price* only once and only upwards and within ten minutes after his/her first bid (states "$S_{c1}$" and "$S_{c2}$" respectively). The BBB App will then automatically relay the *Priority2-Price* to 3-15 drivers that are nearest and have the lowest arrival-time to the customer's location. Each of the 3-15 selected drivers can respond by setting their "preferences" (in the BBB App) to only one of five states: i) to automatically accept the Priority Price (state "$S_{d1}$"); or ii) to bid a percentage of the Priority-Price (eg. 70%-99% of the Priority Price) that exceeds the Standard Price (state "$S_{d2}$"); or iii) to accept the Standard Price (state "$S_{d3}$"); or iv) to manually bid at prices between the Priority Price and the Base Price (state "$S_{d4}$"); or v) not to bid (state "$S_{d5}$") (and collectively, the price bid by driver *i* in the set of drivers 1……..n is $\Omega_{id}$; and the highest price among all bids by all drivers is $\Omega_{cd}$ – also the low price bid by the selected driver with the lowest price-arrival combination). The driver with the highest bid (or a combination of one of the highest bids and lowest arrival-times) that is below the *Priority2-Price* will be automatically selected to provide the ride. Each driver can simultaneously bid for many ride-customer pairs, but any driver that wins a bid cannot bid for any other ride that occurs within 30 minutes after his/her assigned ride. All drivers must inform BBB about when they are off-duty and cannot return to duty unless a specific time elapses; and any driver that doesn't bid for a customer will pay a small fee to BBB. Independent unaffiliated commercial drivers whose vehicles have commercial plates can also bid for Type-5 customers by downloading the BBB App and bidding below the Priority Price. The BBB App will automatically assign a primary-driver (with the lowest bid or a combination of one of the lowest bids and lowest arrival-times) and a secondary driver (with the next highest bid) to the customer to provide the ride. The customer will be notified of a final price before he/she makes payment. All payments will be made through the BBB app. BBB will pay drivers a fee equal to the lesser of: i) the average percentage of revenues that black-car taxi companies pay their drivers in that city less applicable taxes and less any fees payable to third-party car-owners; or ii) X%-Y% (eg. 75%-90%) of the fees paid by the customer less applicable taxes and less any fees payable to third-party car-owners – the variables include commercial insurance fees, road-worthiness testing costs; traffic conditions in the city; the type/class of car used; the customer's gratuities; and the driver's "quality-scores"; etc. (collectively, the fee is "$\psi_{dd}$"). BBB's variable costs and allocated fixed costs for the ride are $B_v$ and $B_f$ respectively. BBB's opportunity cost is the net profit ("$\psi_{ob}$" – the "net" incremental profit over and above that of the matched-ride) from winning and servicing another customer with a probability of $\beta_{ob}$; and the probability $\beta_{lb}$ that another customer with a larger fare exists. The driver's own variable costs and allocated fixed costs for the ride are $D_v$ and $D_f$ respectively. The driver's opportunity cost is the net profit ("$\psi_{od}$" – the "net" incremental profit over and above that of the matched-ride) from bidding for, and winning and servicing another customer with a probability of $\beta_{od}$; and the probability $\beta_{ld}$ that another customer with a larger fare exists. Both probability-adjusted opportunity costs and the two-sided bidding/price-setting protocols incorporate dynamic *Regret Minimization* into the dynamic pricing model (ie. the reduction in *Regret* can be significant and varies with changes in state variables). The probability that there at least one driver will bid is $\beta_{id}$. Note that in most cases, and for each driver-customer pair, $\beta_{ob}$ and $\beta_{od}$ will be different and based on different sets of customers. The probability that driver *i*'s bid wins is $\beta_{dw}$. Generally, it's hoped that $R_{cb} < R_{cd} < R_{cs}$. ∎

**Theorem-5B**: *For All Prices That Are Real Numbers, There Is Least One Equilibrium Payoff That Satisfies The Conditions Of Type-5 Pricing And The Elements Of The Mixed Games (eg. multi-sided auctions; matching games; assignment games; reverse-dutch auctions; multi-stability; and combinatorial auctions) And Dynamic Algorithms Involved; And Can Reduce Deadweight Losses And Antitrust Liability.*
*Proof*:
BBB's objective function and payoff is:
$P_o = \text{Max} [(\min\{\Omega_p; \text{Max}((\Omega_{cd} | \text{Max}(S_{d1};S_{d2};S_{d3};S_{d4};S_{d5})); \Omega_{cs}; 0; \Omega_{cb})\} - (\psi_b*(\beta_{ob} | \beta_{lb})) - (\psi_{dd}+\beta_f+\beta_v))*\beta_{id}]$

The driver's payoff is:
$P_s = \text{Max}[\{(\beta_{id} | \beta_{idw})*(\psi_{dd} - D_v - D_f - (\psi_{od}*(\beta_{od} | \beta_{ld})))\}; 0]$; where $(0<\psi_{dd}) \in \Omega_{cd}$.

These formulas can also answer the question of when a consumer will join a sharing system – and generally and with limited consideration of the customers' non-monetary utility from expedited service and or convenience, a customer is more likely to join the sharing system as $(P_o+P_s) \to -\infty$.

The dynamic pricing model reduces *deadweight losses* because it dynamically reduces both *consumer-surplus* and *supplier-surplus* (ie. the reductions vary with the bidding patterns and other state variables). ∎

**Theorem-6**: *For Each Of Type1, Type-2, Type-3, Type-4 And Type-5 Pricing, There Is Or There Can Be Multi-Stability, But There Is No Unique Equilibrium Or Pure-Nash Equilibrium.*
*Proof*:
There is or there can be *Multi-stability* in each of the Type1, Type-2, Type-3, Type-4 And Type-5 pricing mechanisms and there is no unique equilibrium or pure Nash Equilibrium simply because of any of the following factors: i) the "cross-contingent auction" feature (introduced herein and above); ii) the combinatorial feature of combinatorial auctions; iii) each player cannot know the equilibrium strategies of the other players; iv) in some circumstances, a player can gain by changing only their own strategy (while other players don't change their strategies), but that is not very common. ∎

**Theorem-7**: *Contrary To Generally Accepted Theorems, And For Each Of Type1, Type-2, Type-3, Type-4 And Type-5 Pricing Mechanisms, Any Conditionality (x│y) That Involves Any Of The Probabilities $β_{ob}$, $β_{lb}$, $β_{od}$, $β_{ld}$ And $β_{dw}$, Is "Non-Recursive".*
*Proof*:
Its widely accepted in Math and statistics (especially after Kolmogorov) that conditional probabilities are recursive. Contrary to those theorems, in the Type1, Type-2, Type-3, Type-4 And Type-5 pricing-mechanism payoff functions (defined herein and above), all conditional probabilities are non-recursive because: i) any event that pertains to any provider-customer (driver-customer) pair does not wholly depend on their history of transactions or the history of each person's transactions with similar counter-parties in the system; ii) the "cross-contingent auction" feature of the bidding and price-setting system (explained herein and above) increases the possibility (and almost ensures) that any prior patterns of bidding will not be repeated in the future or that probabilities are recursive because of the changing combinations of participants in both sides of the system; iii) the combinatorial auction feature of the pricing system, significantly eliminates recursion of probabilities and makes it almost unlikely that any prior bidding patterns will be repeated; iv) the bidding by any driver-customer pair can be affected by bidding by another related or un-related driver-customer pair; v) as a result, there is no or very minimal Status Quo Bias. That implies that most or a significant portion of Bayesian statistics is or may be wrong. The issue of conditional probabilities is also relevant to Mckelveya & Page (2002) and Wolitzsky (2016) which analyzed the relationship between *Bayesian Equilibrium* and the *Myerson–Satterthwaite Impossibility Theorem*. Gelman (2008), Thompson (2007) and Cox (2006) critiqued Bayesian statistics. Hahn (2014) noted the increasing use of Bayesian statistics in modern cognition and psychometric studies, and surveyed articles that critiqued Bayesian Statistics. ∎

**Theorem-8**: *Since The Probabilities $β_{ob}$, $β_{lb}$, $β_{od}$, $β_{ld}$ and $β_{dw}$ (For Each Of Type1, Type-2, Type-3, Type-4 And Type-5 Pricing Mechanisms) Are "Non-Recursive Probabilities", The Myerson-Satterthwaite Impossibility Theorem Is Null And Void (Or Does Not apply In all Circumstances).*
*Proof*:
The issue of recursiveness of conditional probabilities is also relevant to Bayesian Equilibrium, which is based on Baye's Rule. Since the *Myerson–Satterthwaite Impossibility Theorem* is partly based on Bayesian Equilibrium, and Bayes Theorem is wrong or does not apply in all circumstances (as shown above) and the Probabilities $β_{ob}$, $β_{lb}$, $β_{od}$, $β_{ld}$ and $β_{dw}$ are "Non-Recursive Probabilities" then the *Myerson-Satterthwaite Impossibility Theorem* is null and void. Mckelveya & Page (2002), Colini-Baldeschi, Goldberg, et. al. (Nov. 2017), Othman & Sandholm (2009) and Wolitzsky (2016) discussed the relationship between *Bayesian Equilibrium* and the *Myerson–Satterthwaite Impossibility Theorem*.

Separately, given the assumptions underlying Bayesian statistics and Bayesian Equilibrium, and Theorems 6 & 7 herein, the *Myerson-Satterthwaite Impossibility Theorem* cannot be valid. Under *Perfect Bayesian Equlibrium*, beliefs and strategies must comply with the following conditions:

i) *Sequential Rationality*: given the beliefs, each strategy should be optimal in expectation. A "belief system" refers to allocation of probabilities to nodes in a game here the sum of all such probabilities is one.
ii) *Consistency:* each belief should be updated according to *Baye's Rule* and the strategies, for all non-zero probabilities (but in paths of zero probability, aka off-the-equilibrium paths, the beliefs can be arbitrary).

In this instance, there cannot be *Sequential Rationality* or *Consistency* because of the factors mentioned above in Theorems 6 & 7. ∎

**Theorem-9**: *The Mechanisms Introduced In Colini-Baldeschi, Goldberg, et. al. (Nov. 2017), Does Not Provide IR (Individual Rationality), IC (Incentive Compatibility) And SBB (strong budget-balance).*
*Proof*:
Colini-Baldeschi, Goldberg, et. al. (Nov. 2017) claimed that their mechanism was the first IR (Individual Rationality), IC (Incentive Compatibility) and SBB (strong budget-balance) mechanisms that provides an O(1)-approximation to the optimal social welfare for two-sided markets. Their claim is false because: i) their mechanisms partly depends on Bayesian Equilibrium which in such instance, cannot hold; and ii) the probabilities implied in their model assume conditionality, which also cannot hold given the above mentioned theorems; iii) their assumption that there is a given probability distribution for sellers' preferences is wrong. ∎

5.7. The New Dynamic Pricing Models Are "*Deep Learning*" Algorithms And They Can Prevent/Reduce *Algorithm Collusion*.

In each of the Type-1, Type-2, Type-3, Type-4 and Type-5 pricing schemes, unforseen events (such as road accidents; fuel shortages), memory (roughly recurring events) and predictable events (such as high road traffic or low customer demand) are accounted for in the dynamic pricing model by: 1) the two-sided auctions for each ride which inherently includes learning by the pricing system, the driver and the customer; 2) "learning" in the calculation of the *Base-Price* by BBB; 3) "learning" in the calculation of the driver's and the customer's opportunity costs.

Thus, the new dynamic pricing models are evolutionary and incorporate *Learning* (ie. beliefs about others' decisions; continuous updating of assumptions and variables; identification of relationships among variables; and *Reinforcement Learning*) because:

1) the models can include new and continuously updated data about state variables (eg. which are used to calculate the *Standard-Price* and *Base-Price*) such as real-life conditions on the road, neighborhood traffic patterns, the driver's mental state; the customer's mental state and preferences; and the state of competition in the ride-sharing market;
2) in some of the models, the auction-protocol involves BBB setting a *Base Price* or *Standard Price* (around which bidding occurs) and then the system "learns" the demand for services around such prices;
3) since the bidding by any driver-customer pair can be affected by bidding by another related or un-related driver-customer pair, each dynamic pricing model implicitly "learns" and incorporates market conditions;
4) as mentioned above, the structure of the dynamic pricing systems can cause both it, and the buyers and sellers (especially those to follow routines) to develop "*long memory*" about bidding patterns, service-experience and state variables such as traffic, weather and other competing companies;
5) each payoff model is essentially a decision model and the probabilities used are akin to decision weights, and thus each such payoff model is stochastic because the subjective probability that an event occurs or that a decision is made can be expressed as the ratio of its own decision-weight to the sum of all weights. See the comments in Mohri & Medina (2014) and Nazerzadeh, Leme, et. al. (2016).

Furthermore, the new pricing algorithms introduced in this article can prevent and or reduce all types of Algorithm Collusion because: i) most of the data sources for prices are "within-system"; ii) updating is done in ways that reduce correlation with other pricing systems; iii) a substantial percentage of the data is "transaction-specific" (tailored to a specific transaction); iv) the company's own cost-structure and operating conditions are given substantial weights in the pricing algorithms; v) the algorithms are structurally different from traditional pricing algorithms.

On *Deep Learning* and *Regret Minimization* in Mechanism Design, see: Feng, Narasimhan & Parkes (2018), Golowich, Narasimhan & Parkes (2018), Dutting, Feng, et. al. (2019), Shen, Peng, et. al. (2017), and Sandholm & Likhodedov (2015). However, many "*Learning*" models suffer from the problems/weaknesses that were explained in Nwogugu (2013b) (invalidity of the mean variance framework), Nwogugu (2017a;b;c;d) (biases in returns; *Framing Effects*; asset pricing errors; invalidity of risk aversion and loss aversion; etc.) and Nwogugu (2007; 2006) (critique of bankruptcy prediction models, LOGIT/PROBIT/regressions, Neural Networks, ARCH/GARCH/ARMA/SV models).

**Theorem-10**: *Each of The New Dynamic Pricing Models Violates The Myerson-Satterthwaite Impossibility Theorem.*
*Proof*:
Each of the five new dynamic pricing models introduced herein has all the following properties:
  i) *Individual Rationality* of buyers - for each buyer, there is a bid such that the buyer always obtains non-negative utility.
  ii) *Individual Rationality* of sellers.
  iii) *Strong Budget Balance* – not all buyer's payments are given to the seller, but the portion given to the seller reflects fair value for services rendered or goods provided.
  iv) *Dominant-Strategy Incentive Compatibility of buyers and sellers* – each buyer and seller can achieve the best outcome to him/herself by reporting/acting according to his/her true preferences/valuations.
  v) *Pareto Optimality* - there is no other allocation/matching superior to that provided by the mechanism in all aspects, provided all buyers report their true valuation (its assumed that mechanism structure encourages truth-telling by buyers and sellers).
  vi) *Feasibility* – with respect to real-world constraints.
  vii) *Privacy* – the identities of bidders (on both sides of the system) are not disclosed and for each "contract" or transaction (eg. a taxi-ride or Airbnb rental), only the winning seller-buyer pair are introduced to each other.
  viii) *Welfare Maximization* (both *Exact And Approximate*) – with regards to the overall system.

Its noteworthy that each of the five new dynamic pricing models violates the *Myerson-Satterthwaite Impossibility Theorem* because each has the characteristics of *individual rationality*; *strong budget balance*; *incentive compatibility* of buyers/sellers and *pareto-optimality* and "learning".
  Mckelveya & Page (2002) and Wolitzsky (2016) analyzed the relationship between *Bayesian Equilibrium* and the *Myerson–Satterthwaite Impossibility Theorem*.

6. A Legal Business Model In Place Of Uber And Lyft.
  The following is a legal business model for a hypothetical company named "BBB" that achieves same objectives as Uber or Lyft or Grab (as of mid-2017, the combined equity market-values of the top-five taxi SEOs such as Uber, Lyft, Grab and Didi Chuxing exceeded US$140 billion). BBB will obtain a blanket taxi license from the city government which will: i) enable BBB's taxis to provide pre-booked rides within the city/town; ii) allow BBB to operate using vehicles that have commercial-plates and commercial licenses; iii) allow BBB to operate using private vehicles that have current private plates, have passed city-sanctioned road-worthiness tests and have valid commercial insurance policies provided by BBB; and that are operated by drivers with ordinary (non-commercial) or commercial driving licenses where each such non-commercial driver will have to obtain a medical certificate from a doctor to confirm that he/she is fit to drive with passengers; iv) require that BBB obtain commercial vehicle insurance for each such driver-vehicle pair or submit copies of the driver's commercial vehicle insurance; v) require that BBB collect tax revenues from independent contractor drivers and affiliated car owners and remit same to the city/state/federal governments. The commercial insurance policies must provide coverage for commercial passengers and non-commercial drivers that carry passengers in commercial transactions. BBB will use the following classes of cars:
  i) BBB will purchase cars or lease cars and obtain commercial plates and commercial vehicle insurance for each such car (Type-A cars). BBB can obtain significant price discounts by buying or leasing large quantities of used cars at the same time and obtaining group insurance and group road-worthiness testing for such cars. BBB can also negotiate with the city government to provide discounts for commercial plate fees and road-worthiness fees (that are administered by the city government). BBB will lease its Type-A cars to un-affiliated drivers on a weekly or monthly basis and BBB will deduct such lease payments and income taxes from revenues generated by each Type-A drivers. The drivers can be non-commercial or commercial drivers. BBB will provide commercial vehicle insurance policies for such non-commercial drivers.

ii) BBB will also enter into strategic alliances with third-party owners of cars that have commercial plates and commercial insurance (Type-B cars) to lease their cars to both commercial and non-commercial drivers on a weekly or monthly basis and such drivers will work within BBB's network. BBB will deduct such lease payments from revenues generated by each Type-B driver, and remit same to the car-owner. BBB will provide commercial vehicle insurance policies for such non-commercial drivers. Each such non-commercial driver will have to obtain a medical certificate from a doctor to confirm that he/she is fit to drive with passengers.

iii) BBB will enter into strategic alliances with third-party drivers that have commercial driver's licenses and own commercial cars which will have commercial plates and commercial insurance (Type-C cars) to drive their cars for hire within the BBB network.

iv) BBB will enter into strategic alliances with third-party drivers that own private cars that have private plates to provide rides within the BBB network (Type-D cars). BBB will obtain and pay for commercial vehicle insurance (purchased or self insurance) for each such car/driver; and commercial-quality road-worthiness tests for each such car. Each such non-commercial driver will have to obtain a medical certificate from a doctor to confirm that he/she is fit to drive passengers.

v) For all four classes of cars, "*commercial insurance for private cars and or non-commercial drivers*" means that the commercial insurance policy adjusts for, and covers the specific risks that: i) the driver is not a professional driver, may be impaired (eg. eyesight), and is less skilled than commercial drivers, and may cause more accidents; ii) there are passengers in the vehicle who are subject to accidents and other harm (eg. defective car seats and or seat-belts and or air-conditioners; un-sanitary conditions in the car; etc.); iii) the non-commercial driver may not always ensure compliance with, and the private car may not always comply with all standards/requirements that apply to commercial vehicles (such as road-worthiness tests; emissions; etc.). *Self insurance* means that BBB will create a subsidiary that will provide the self insurance and create balance sheet cash reserves for such risk; and pay-out funds when claims are made; and where feasible, re-insure the risk. The balance sheet reserves must be determined by, and satisfactory to the city/state government and insurance regulators and reviewed at least quarterly.

Each Type-A, Type-C, Type-D and Type-B car will be fitted with BBB's sensors which will be linked to the BBB-App and the "external GPS" and will detect the duration-of-occupancy of each passenger in the back seat of such car, and help in calculating an "estimated revenue" for each such passenger (based on estimated distance per minute, meter rates and the customer's duration of occupancy) which will be compared to the official "actual revenues" generated by that car through the BBB app. Each Type-A, Type-C, Type-D and Type-B car will be fitted with an external GPS tracking system (separate from the GPS system in the BBB App). All drivers will function as independent consultants. BBB will deduct taxes payable by each independent contractor driver and remit same to the city, state and federal governments. BBB will develop its ride-sharing app (which will contain a GPS system) which prospective drivers, customers and owners of cars can download onto their own cellphones (BBB will not provide cellphones to drivers or car owners). The BBB app will be linked to the sensors such that BBB and an owner of the Type-B car can track the driver's location and ride-statistics (eg. number of rides for each week; revenues; estimated fuel consumption; etc.) in real time.

7. Conclusion.

There are several characteristics of the *Sharing Economy* that make traditional theoretical computer science and economic theory hard to apply and some are: i) multi-sided systems/markets; ii) transferable utility.

The new dynamic pricing models introduced herein address current and future problems inherent in pricing mechanisms in the sharing economy and can be used by other non-taxi SEO companies – such as Airbnb, WeWork and Liquidspace.

**Disclosures:**
Funding: This study wasn't funded by any grant.

Conflict of Interest: I have no conflict of interest.